\documentclass[letterpaper]{jpconf}
\usepackage{graphicx}
\usepackage{epsf}
\usepackage{psfig}
\def\plottwo#1#2{\centering \leavevmode
\epsfxsize=.45\textwidth \epsfbox{#1} \hfil
\epsfxsize=.45\textwidth \epsfbox{#2}}
\def\apj{ApJ}
\def\apjs{ApJS}

\def\apjl{ApJL}

\def\pre{Phys. Rev. E}

\def\araa{ARA\&A}
\def\mnras{MNRAS}
\def\jcp{J. Comp. Phys.}

\begin{document}
\title{Simulating Supersonic Turbulence 
in Magnetized Molecular Clouds}

\author{Alexei G Kritsuk$^1$, Sergey D Ustyugov$^{2,1}$, Michael L Norman$^1$ and Paolo Padoan$^1$}

\address{$^1$ University of California - San Diego, 9500 Gilman Drive, Mail Code 0424, La Jolla, 
         CA 92093-0424, USA}
\address{$^2$ Keldysh Institute of Applied Mathematics, Russian Academy of Sciences, Miusskaya Sq. 4,
Moscow 125047, RUSSIA}

\ead{akritsuk@ucsd.edu}

\begin{abstract}
We present results of large-scale three-dimensional simulations of weakly magnetized supersonic 
turbulence at grid resolutions up to $1024^3$ cells. Our numerical experiments are carried out
with the Piecewise Parabolic Method on a Local Stencil and assume an isothermal equation of state.
The turbulence is driven by a large-scale isotropic solenoidal force in a periodic computational domain
and fully develops in a few flow crossing times. We then evolve the flow for a number of flow 
crossing times and analyze various statistical properties of the saturated turbulent state. We show
that the energy transfer rate in the inertial range of scales is surprisingly close to a constant,
indicating that Kolmogorov's phenomenology for incompressible turbulence can be extended to
magnetized supersonic flows. We also discuss numerical dissipation effects and convergence of
different turbulence diagnostics as grid resolution refines from $256^3$ to $1024^3$ cells. 
\end{abstract}

\section{Introduction}
Supersonic turbulence plays an important role in shaping hierarchical internal 
substructure of molecular clouds (MCs) \cite{elmegreen.04,scalo.04}. This turbulence 
is the key ingredient in statistical theories of star formation as it controls
initial conditions for gravitationally collapsing objects \cite{padoan.02,mckee.07,padoan....07}. 
The nature of molecular cloud turbulence is still poorly understood and no 
simple conceptual theory exists akin to that of Kolmogorov's phenomenology for 
turbulence in fluids. Two major complications are {\em strong compressibility}
of interstellar gas mediated by radiative energy losses and presence of dynamically 
important {\em magnetic fields}. While these two factors bring into play additional 
variables and equations rendering the modeling hardly tractable analytically and 
complicating numerical analysis, the nature of underlying nonlinearity 
contained in the advection terms remains the same. From a mathematician's
perspective, the general problem is essentially the same as in the Navier-Stokes 
turbulence, one of the remaining six unsolved Millennium Prize Problems for which 
the Clay Institute of Mathematics is offering a bonus of \$1 million.\footnote{\tt 
http://www.claymath.org/millennium/Navier-Stokes\_Equations/}
Thus hopes to find universality in interstellar turbulence are not entirely
unfounded, see e.g. \cite{heyer.04}, although some researchers believe that
breakdown of universality in incompressible isotropic magnetohydrodynamic (MHD) 
turbulence is possible, e.g. due to nonlocality of interactions \cite{yousef..07,beresnyak.09}.

The supersonic regime typical of MC turbulence is extremely hard to achieve in the 
laboratory and the information available from astronomical observations 
is rather limited \cite{heyer.04}. This makes numerical simulations, perhaps, the 
only tool available to explore the statistics of supersonic turbulence in detail. 
High dynamic range simulations can shed light on the energy transfer between scales and 
on the key spatial correlations of the relevant fields in these flows only if they
provide sufficient scale separation to resolve the dynamics of prevailing nonlinear 
interactions in the inertial range. In addition, low numerical dissipation and 
high-order numerical methods for three-dimensional compressible MHD are needed to 
achieve an accurate description of the inertial range of supersonic turbulence. 

Stable, accurate, divergence-free simulation of magnetized supersonic turbulence is a severe 
test of numerical MHD schemes and has been surprisingly difficult to achieve due to the range 
of flow conditions present. Over the last two years we have developed Piecewise Parabolic Method 
on a Local Stencil (PPML), a new higher order-accurate, low dissipation numerical method which 
requires no additional dissipation or local ``fixes'' for stable execution \cite{popov.07,popov.08,ustyugov...09}. 
PPML is a local stencil variant of the popular PPM algorithm \cite{colella.84} for solving the 
equations of compressible ideal magnetohydrodynamics. The principal difference between PPML and 
PPM is that cell interface states are ``evolved'' rather that reconstructed at every timestep, 
resulting in a compact stencil. Interface states are evolved using Riemann invariants containing 
all transverse derivative information. The scheme is fully multidimensional as the conservation 
laws are updated in an unsplit fashion, including the terms corresponding to the tangential 
directions in the amplitude equations. This helps to avoid numerous well-known pathologies, such 
as ``carbuncles'', etc. \cite{quirk94}. Divergence-free evolution of the magnetic field is 
maintained using the higher order-accurate constrained transport technique \cite{gardiner.05}. 
The low dissipation and wide spectral bandwidth of this method make it an ideal choice for 
direct simulations of compressible turbulence.

In Ref. \cite{kritsuk...09} we reported results from a set of three simulations of driven 
isothermal supersonic turbulence at a sonic Mach number of 10 on $512^3$ meshes 
demonstrating the performance of our PPML solver on models with different degrees of
magnetization from super-Alfv\'enic through trans-Alfv\'enic regimes. Turbulence 
statistics derived from these three-dimensional models show that the density, velocity, 
and magnetic energy spectra vary strongly with the strength of turbulent magnetic field 
fluctuations. At the same time, the three cases suggest that a linear scaling in the 4/3-law
of incompressible MHD turbulence \cite{politano.98a,politano.98b} is preserved even in
the strongly compressible regime at $M_s=10$, if the proper density weighting is applied to
the Els\"asser fields following the 1/3-rule introduced in 
Refs.~\cite{kritsuk...09,kritsuk...07a,kritsuk...07b,kowal.07}, see Section~\ref{4.2} for more detail.

In this paper we present results from PPML simulations of statistically isotropic MHD turbulence 
at a sonic Mach number $M_s=10$ and Alfv\'enic Mach number $M_A=3$ on grids from $256^3$ to 
$1024^3$ cells. We use this set of simulations to check the (self-)convergence rates for 
various statistical measures with improved grid resolution. We also confirm the universal
linear scaling for the mixed third-order structure functions of modified Els\"asser fields 
discussed earlier in \cite{kritsuk...09}.

\section{Numerical Model and Parameters}
We use PPML to solve numerically the MHD equations for an ideal isothermal gas in a cubic 
domain of size $L=1$ with periodic boundary conditions
\begin{equation}
\frac{\partial \rho}{\partial t}+{\bf \nabla}\cdot(\rho {\bf u}) =0,
\end{equation}
\begin{equation}
\frac{\partial \rho {\bf u}}{\partial t}+{\bf \nabla\cdot}\left[\rho {\bf uu} -
{\bf BB}+ \left(p+\frac{{\bf B}^2}{2}\right){\bf I}\right]={\bf F},\label{mome}
\end{equation}
\begin{equation}
\frac{\partial {\bf B}}{\partial t}+{\bf \nabla}\cdot({\bf uB} - {\bf Bu}) = 0.\label{fara}
\end{equation}
Here $\rho$ and ${\bf u}$ are the gas density and velocity, ${\bf B}$ is the magnetic field, 
the gas pressure $p\equiv c_s^2\rho$ and the isothermal sound speed $c_s\equiv1$, ${\bf I}$ is
the unit tensor. 
PPML conserves mass, momentum and magnetic flux, and keeps ${\bf \nabla\cdot B}=0$ to the 
machine precision. A fixed large-scale ($k\le2$) isotropic nonhelical solenoidal force 
${\bf F}\equiv\rho{\bf a}-\left<\rho{\bf a}\right>$, where ${\bf a}$ is the acceleration
and $\left<\ldots\right>$ indicates averaging over the entire computational domain,
is used to stir the gas keeping $M_s$ 
close to 10 during the simulation. The models are initiated with a uniform density
$\rho_0\equiv1$, large-scale velocity field ${\bf u}_0=\tau{\bf a}$, where $\tau$ is a constant such that 
$\left<{\bf u}_0^2\right>^{1/2}=8$, and a uniform magnetic field ${\bf B_0}$ aligned with the $x$-coordinate 
direction. The magnetic field strength is parameterized by the ratio of thermal-to-magnetic 
pressure $\beta_0\equiv p_0/{\bf B_0}^2$. In Ref.~\cite{kritsuk...09} we carried out three pilot
simulations with $\beta_0=0.2$, 2, and 20 at $512^3$. Here we present results for three 
simulations with $\beta_0=2$ and grid resolutions $256^3$, $512^3$, and $1024^3$ cells.
The largest of these three experiments was carried out using 2048 cores of {\em Ranger} 
supercomputer at TACC. 

Our numerical experiments covered the transition to turbulence and evolution for up to eight dynamical times, 
$t_d\equiv L/2M_s$. 
The magnetic field was not forced and receives energy passively through 
interaction with the velocity field. This random forcing does not generate a mean field, 
but still leads to amplification of the small-scale magnetic energy through compression in shocks 
in combination with flux freezing and, possibly to some extent, via a process known as small-scale 
dynamo, e.g. \cite{schekochihin.07} and references therein.
It is implicitly assumed that the effective magnetic Prandtl number $P_M\approx1$ 
in our numerical models. This is far from the realistic values for the interstellar conditions, 
where  $P_M\gg1$. The limitations imposed by the available computational resources allow us to
reach effective Reynolds numbers on the order of $10^4$ in our largest MHD simulations, while the 
realistic values for molecular clouds can be as large as $\sim10^8$.

\section{The Saturated Turbulent State and Convergence}
\begin{figure}[th]\centering
\includegraphics[width=34pc]{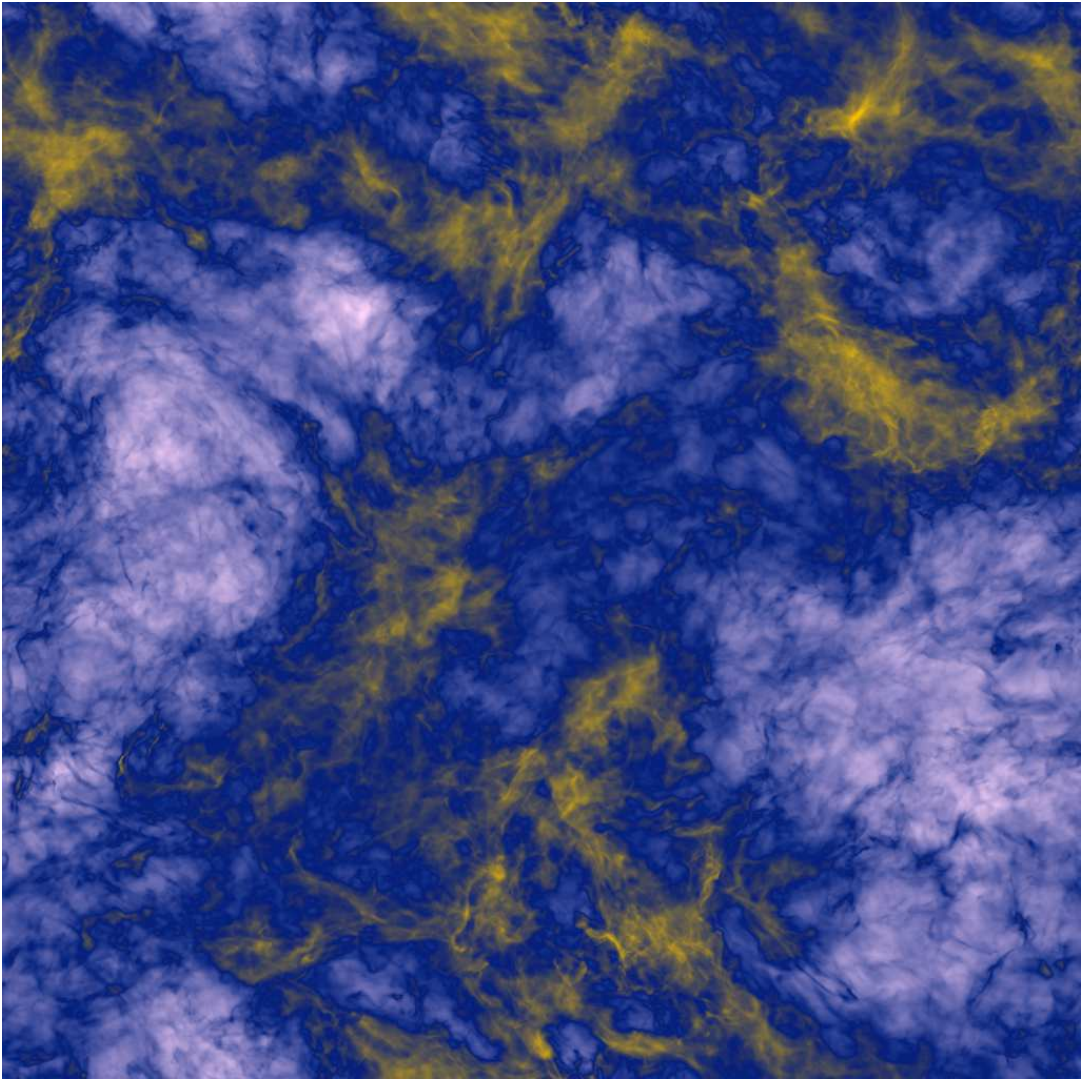}
\caption{\label{project} Projected (column) density image for a snapshot at $t=6.35t_d$
from a $1024^3$ PPML simulation of MHD turbulence at $M_s=10$ and $M_A=3$ carried out on
{\em Ranger} (TACC) using 2048 cores. White-blue-yellow colors correspond to 
low-intermediate-high projected density values. The dynamic range of the image is about 100.
}
\end{figure}

After the initial stir-up phase, our simulations reach a statistical steady state.
A snapshot of the projected (column) density in such a saturated state is shown in Fig.~\ref{project}.
The morphology of structures visible in this projected density distribution is surprisingly
similar to that from a non-magnetized simulation at $M_s=6$ presented in Fig.~1. of 
Ref.~\cite{kritsuk...09}, even though the power spectrum index is about $-1.7$ here 
(see Section~\ref{scal}.1 and Fig.~\ref{figfour}{\bf d}) and $-2.0$ there 
(see Fig.~2{\bf a} in Ref.~\cite{kritsuk...09}) and the
morphology of the density distribution in thin slices (not shown) is clearly different. This 
illustrates the loss of information in the projection procedure that makes it difficult
to measure physical parameters of supersonic turbulent flows from the observed projected density 
distributions alone.

The properties of this saturated state representing fully developed macroscopically isotropic 
compressible MHD turbulence are the main focus of this paper. While some level of physical 
understanding of small-scale rms field amplification and its saturation exists, the structure 
of the saturated state is poorly known, even in the incompressible case \cite{yousef..07}.

\begin{figure}
\centering
\plottwo{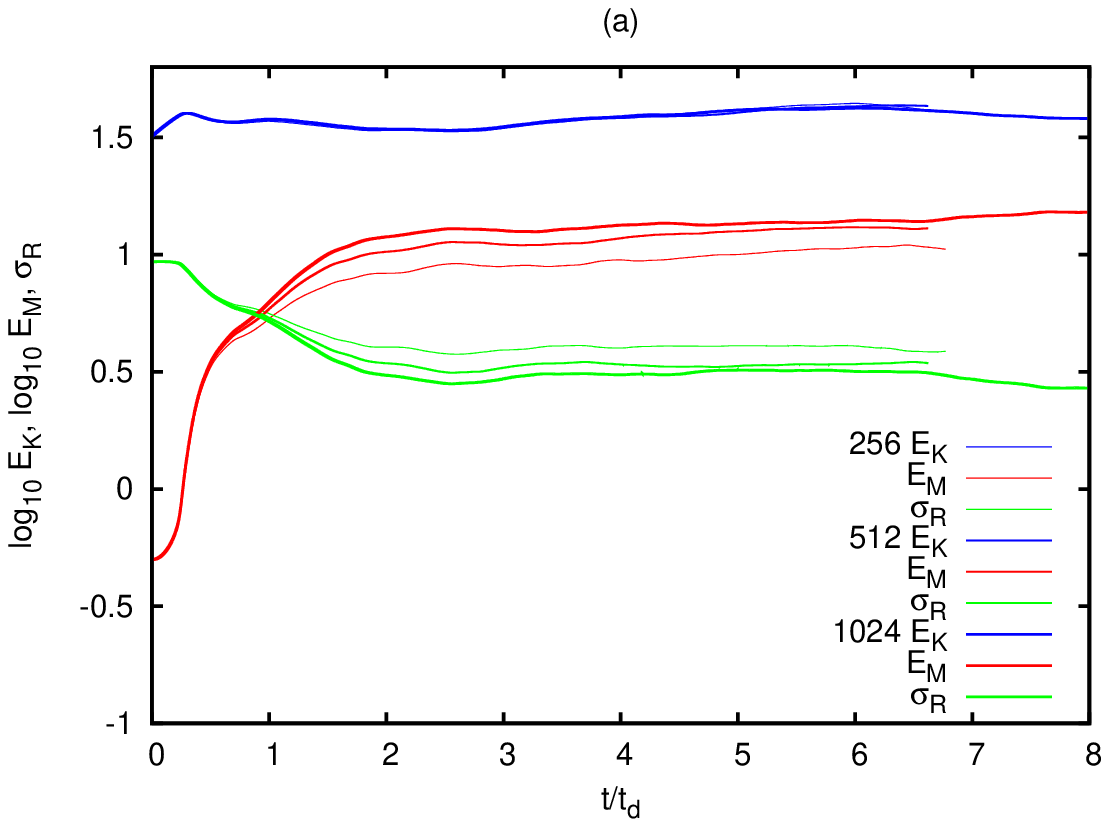}{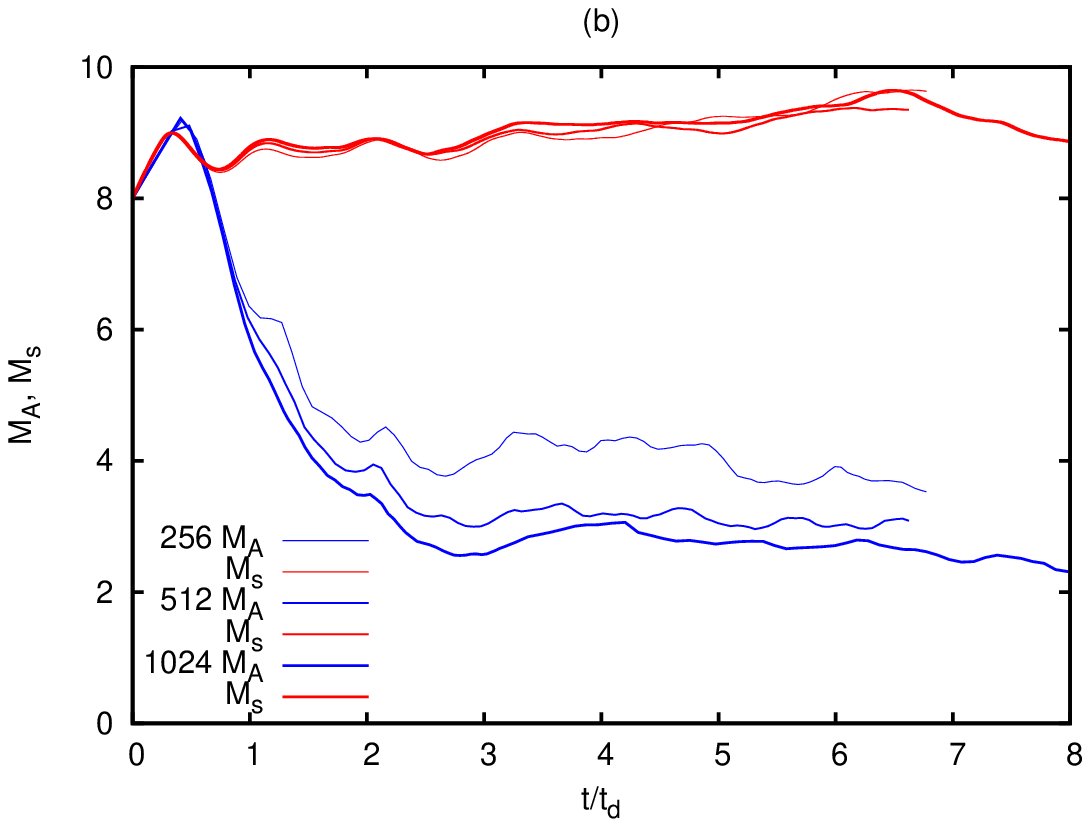}
\plottwo{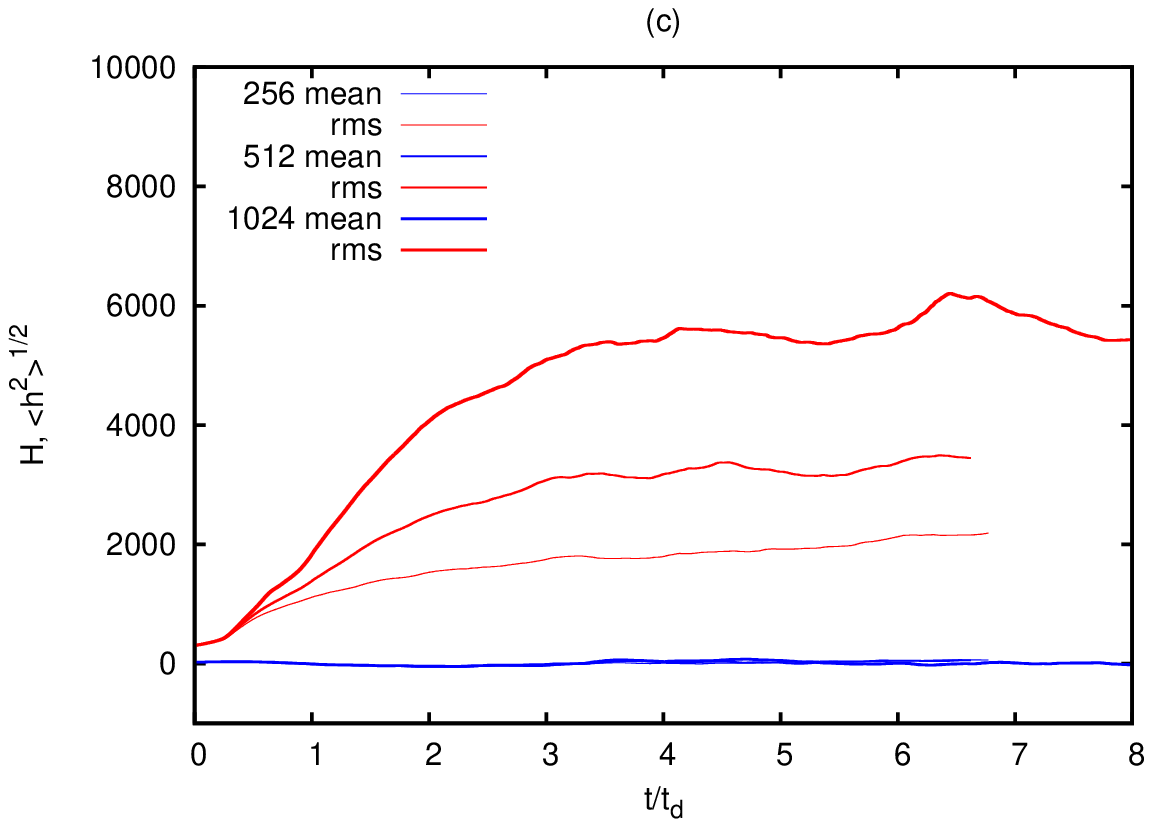}{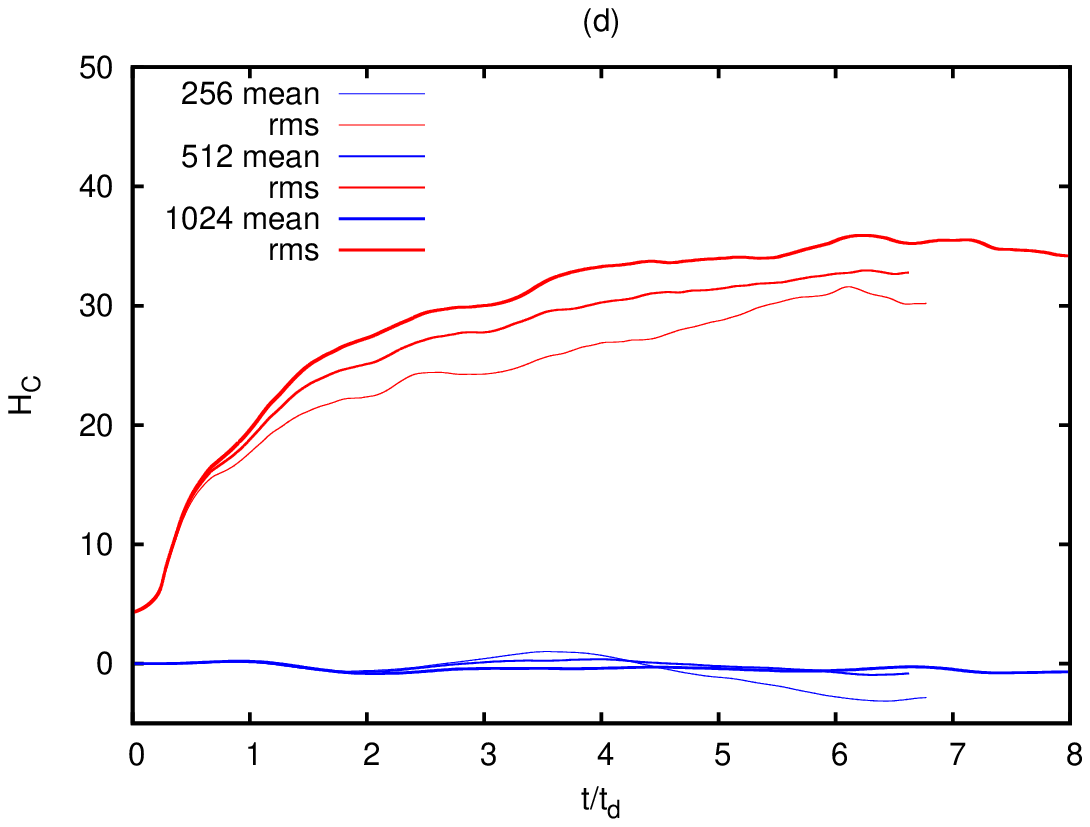}
\plottwo{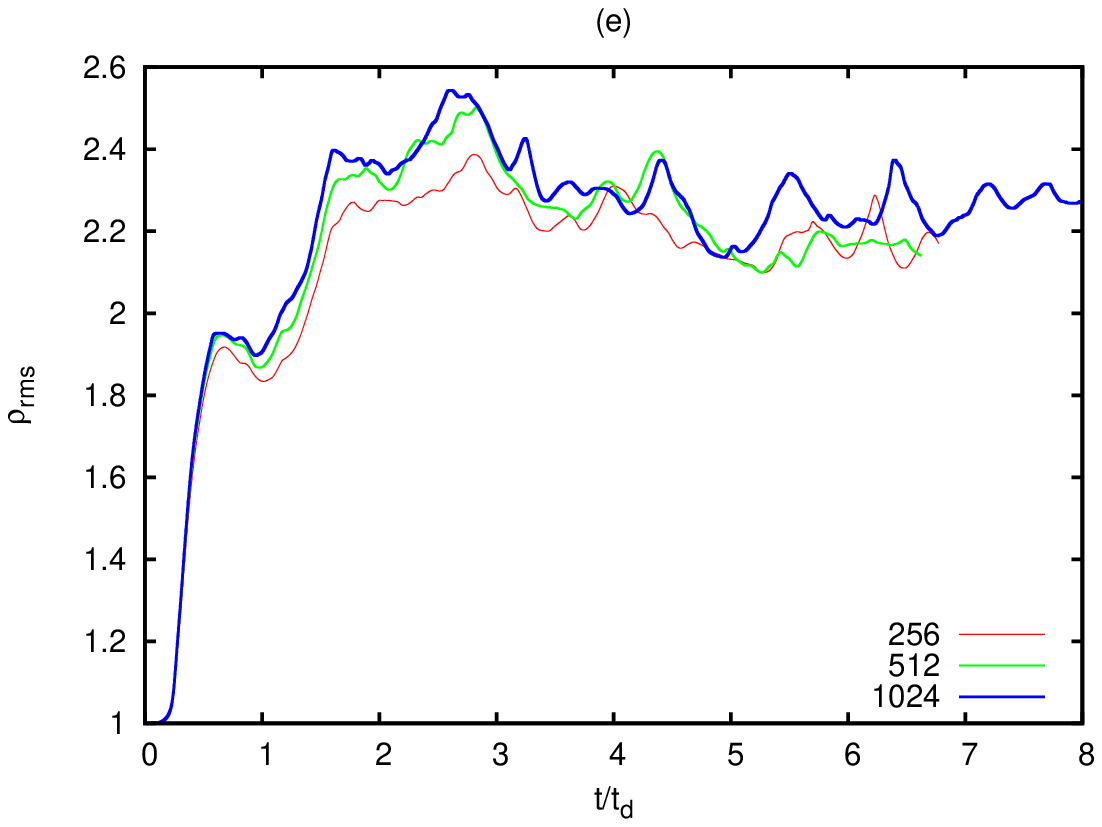}{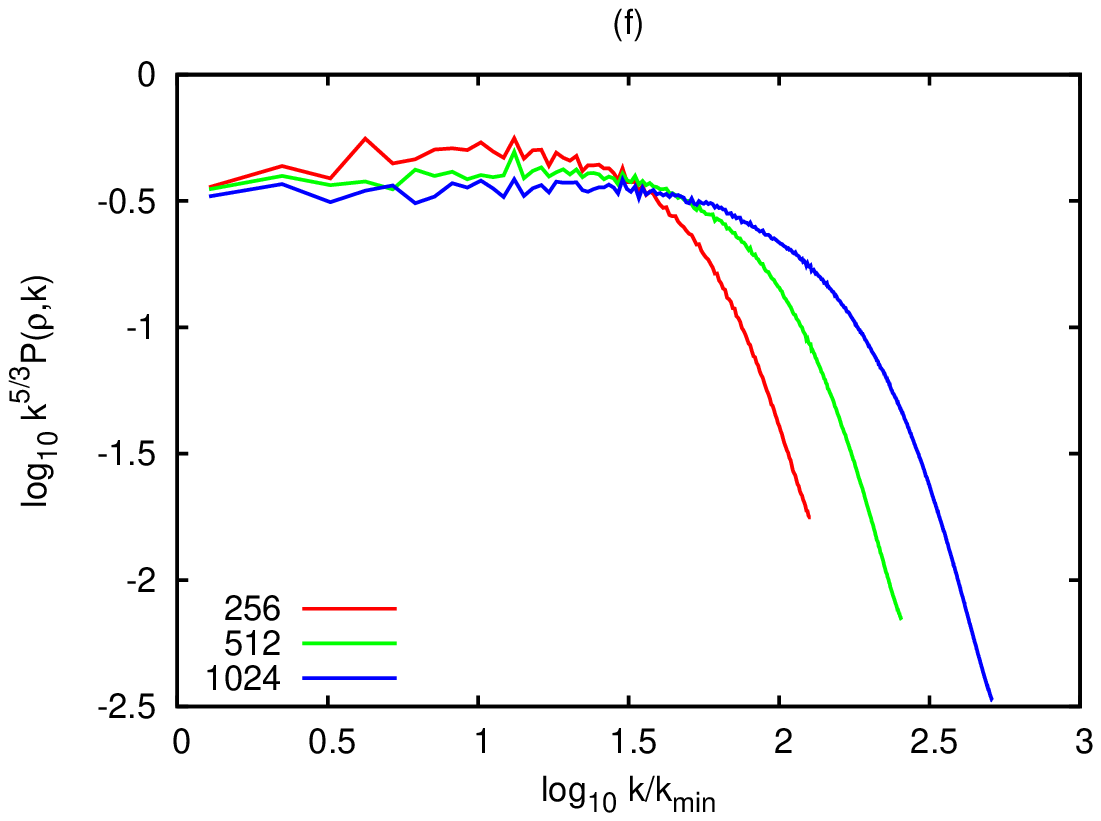}
\plottwo{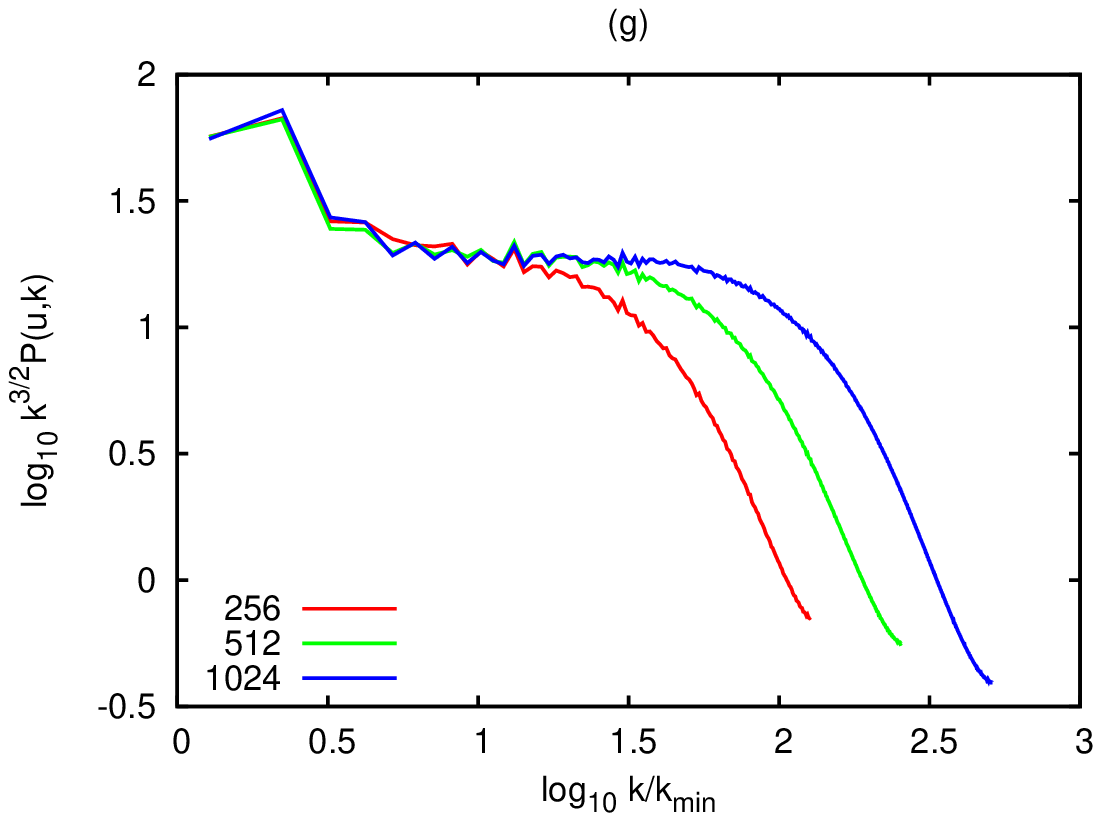}{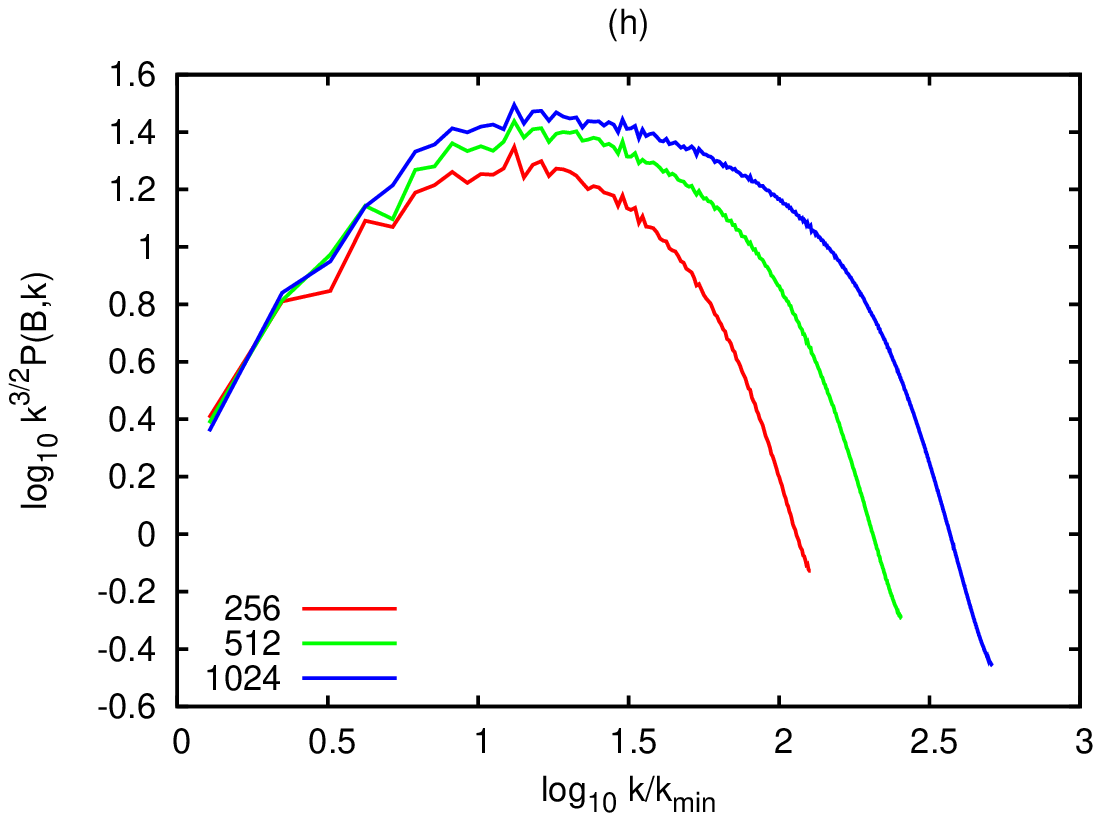}

\caption{\footnotesize Statistics of MHD turbulence at $M_s=10$ and $\beta_0=2$, from PPML 
simulations at grid resolution of $256^3$, $512^3$, and $1024^3$ cells: 
({\bf a}) mean kinetic, magnetic, and residual energy ($\sigma_R\equiv(E_K-E_M)/E_T$) vs. time;
({\bf b}) rms sonic and Alfv\'enic Mach numbers;
({\bf c}) mean and rms helicity;
({\bf d}) mean and rms cross-helicity;
({\bf e}) rms gas density;
({\bf f}) compensated density power spectra at $t=3.5t_d$;
({\bf g}) compensated velocity power spectra at $t=3.5t_d$;
({\bf h}) compensated magnetic energy power spectra at $t=3.5t_d$.
}
\label{figthree}
\end{figure}

We compared the saturation levels for several integral characteristics of turbulence for
models with three different resolutions: $256^3$, $512^3$, and $1024^3$. Since in the $\beta_0=2$
case the kinetic energy $E_K\equiv\int\frac{1}{2}\rho u^2dV$ dominates over the magnetic energy 
$E_M\equiv\int\frac{1}{2} B^2dV$ by a factor of about 3,
it is practical to consider $E_K$ and $E_M$ separately instead of concentrating on the total
energy $E_T=E_K+E_M$. Figure~\ref{figthree}{\bf a} shows that the kinetic energy is already 
converged on $256^3$ grids, in agreement with \cite{lemaster.09}. In contrast, the saturation
level of magnetic energy at $\beta_0=2$ clearly depends on the grid resolution and there are
good chances that convergence can be reached already at resolution of $2048^3$ cells. The 
mean values of $E_M$ for $t\in[3,7]$ are 10.0, 12.3, and 13.5 with the highest corresponding 
to our finest grid resolution of $1024^3$. The lack of convergence in $E_M$ stems from 
numerical diffusion suppressing small-scale compressions and also the small-scale dynamo 
action, if any, in low-resolution runs. The mean value of $E_K$ for the same averaging 
interval is 39.8, thus the magnetic energy contributes about one quarter of the total energy, 
$E_M/E_T\approx0.25$, in the $1024^3$ model.

While the root mean square (rms) sonic Mach number $M_s$ saturates at a level of about
9.2 independent of the grid resolution, the Alfv\'en Mach number $M_A\approx4.0$, 3.1, 
and 2.8 for grids with $256^3$, $512^3$, and $1024^3$ cells, respectively
(see Fig.~\ref{figthree}{\bf b}). 
With the same initial uniform field ${\bf B}_0$, we get turbulent states with
slightly different $M_A$ values with a tendency towards less super-Alfv\'enic flows at
higher grid resolutions.

Since we do not include the explicit dissipative terms in the equations we integrate
numerically, the dissipation in our models is purely numerical and the higher resolution models 
correspond to higher effective Reynolds numbers.\footnote{We follow here the approach developed
in Ref.~\cite{sytine....00} for nonmagnetized simulations with PPM and assume that the nature
of dissipation (purely numerical in our case) does not affect the dynamics in the inertial range 
of scales. Resolving the explicit dissipative terms in Eqs.~(\ref{mome}) and (\ref{fara}) would 
require a much higher grid resolution to obtain an equivalent scale separation.} 
This can be traced in the non-converging levels of rms helicity gradually growing with higher grid 
resolution, see  Fig.~\ref{figthree}{\bf c}.
Since our driving force is non-helical to a very good approximation, the mean helicity 
stays close to zero over the course of a simulation. Figure~\ref{figthree}{\bf d} shows
a rather slow convergence in the levels of the rms cross-helicity and finite deviations
of the mean cross-helicity $H_C\equiv\int{\bf u\cdot B}dV$ from zero in low resolution
runs. PPML does not exactly conserve cross-helicity, even though it is one of the
ideal invariants in MHD. The normalized mean cross-helicity is, however, contained within 
$\pm0.8$\% for $\beta_0=2$ at $1024^3$. Even though both the kinetic and the magnetic
energy as well as the rms Alfv\'en Mach number reach saturation after about 3 dynamical 
times of evolution in our models with $\beta_0=2$, the rms cross-helicity continues 
to grow slowly for up to 4 or even 5 dynamical times meaning that stirring the flow for 
$3t_d$ might not be sufficient to reach a statistical steady state.

Another global diagnostic to look at is the rms density. In models with higher resolution
the flow is better resolved and higher level of density fluctuations is observed. The
rms denstity is, however, not very sensitive to the grid resolution and changes from 2.2 to 
2.3 as the resolution jumps from $256^3$ to $1024^3$, see Fig.~\ref{figthree}{\bf e}.

Finally, to illustrate the convergence of spectral properties of turbulent
fluctuations, in Fig.~\ref{figthree}{\bf f, g}, and {\bf h} we show compensated power
spectra of the density, the velocity, and the magnetic field for a set of snapshots
from simulations with different resolutions taken early in the saturated state at
$t=3.5t_d$. The velocity spectra can be used to access the spectral bandwidth of PPML.
As one can see from Fig.~\ref{figthree}{\bf g}, the velocity spectrum is resolved up
to $\log_{10}k/k_{min}\approx1.1$ at $256^3$ and up to $\log_{10}k/k_{min}\approx1.4$ 
at $512^3$, consistent with the grid refinement by a factor of two.

\section{Exploring Scaling Laws in Supersonic MHD Turbulence\label{scal}}
Here we present time-averaged statistics for our highest resolution $\beta_0=2$ model at 
$1024^3$ grid cells and compare these with our previous results for non-magnetized (HD) 
simulations at $M_s=6$ and grid resolutions of $1024^3$ and $2048^3$ cells 
\cite{kritsuk...07a,kritsuk...07b,kritsuk...09}. Our MHD data are based on 100 full data snapshots 
equally spaced in time for $t\in[3,8]t_d$, i.e. we do the averaging over 
five dynamical times which makes it, perhaps, the best available sample for compressible 
MHD turbulence at this resolution. Our projected density spectra are based on a sample of 
750 projections, including all three projection directions, covering the same time 
span $t\in[3,8]t_d$. We first begin with the scaling properties of the basic fields,
such as the density, velocity, and magnetic field and then discuss ideal invariants,
such as the total energy and the energy transfer rate that may or may not display
universal behavior.

\subsection{Basic Fields}

\begin{figure}
\centering
\plottwo{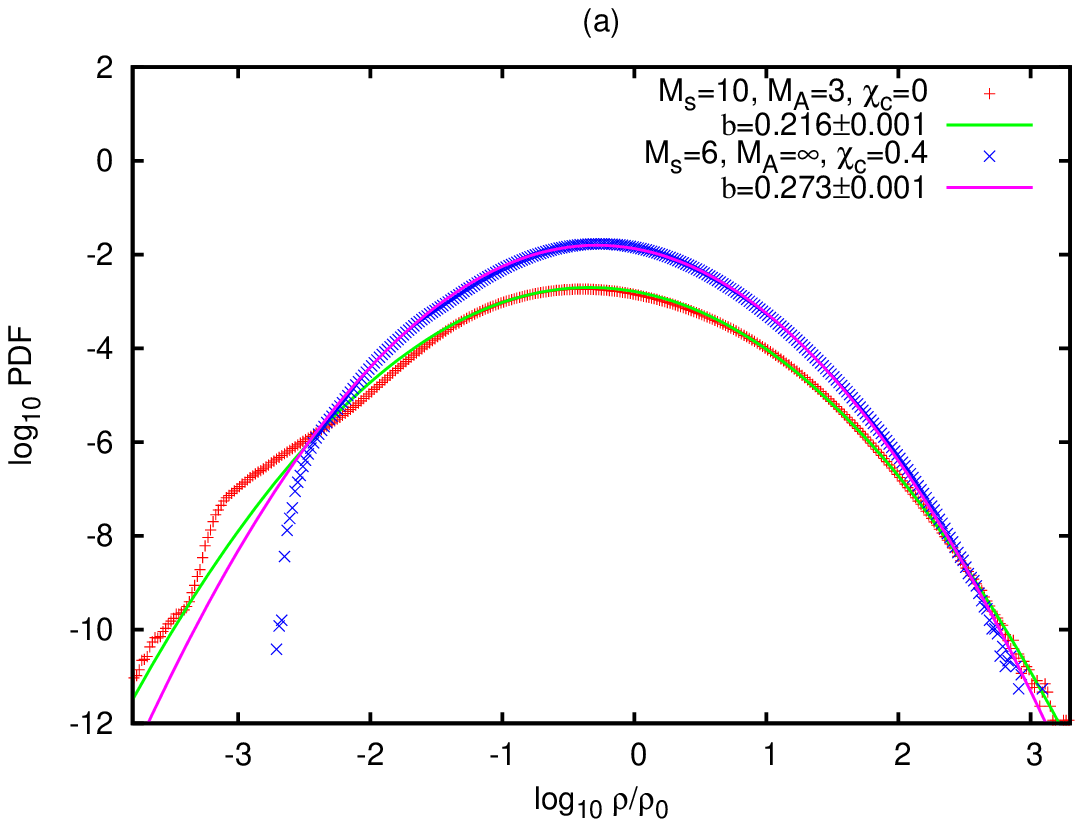}{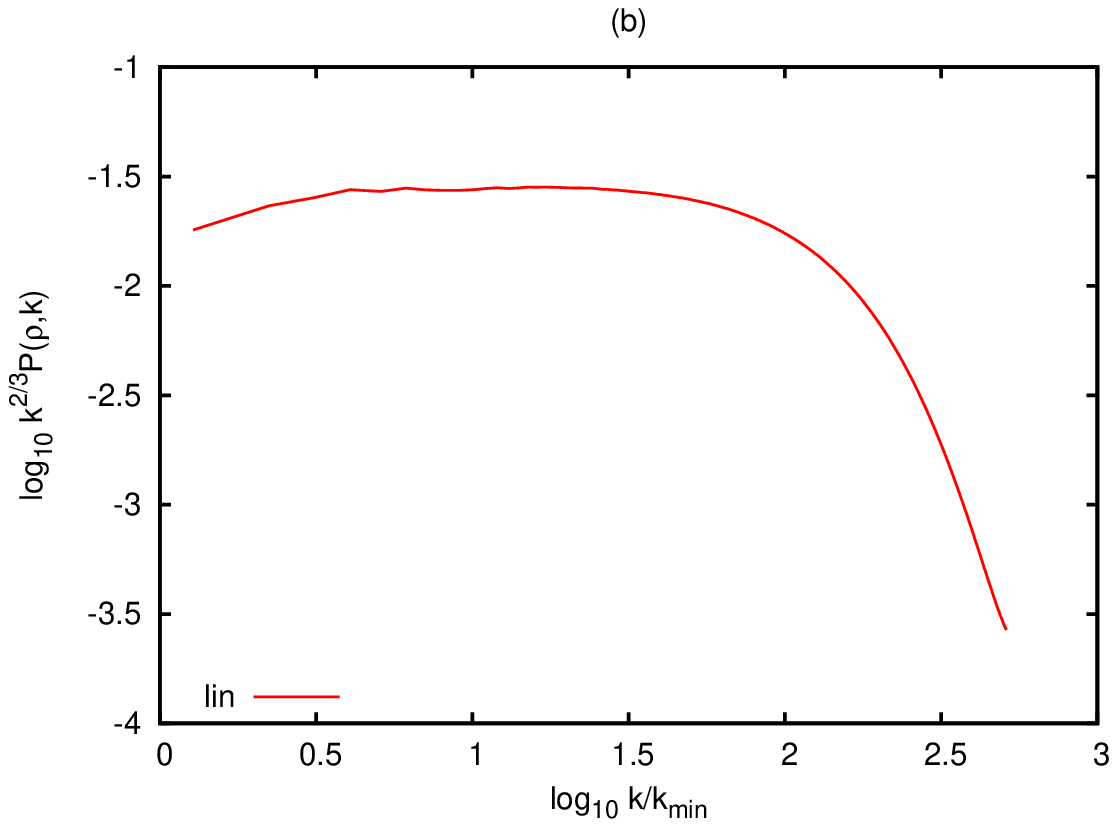}
\plottwo{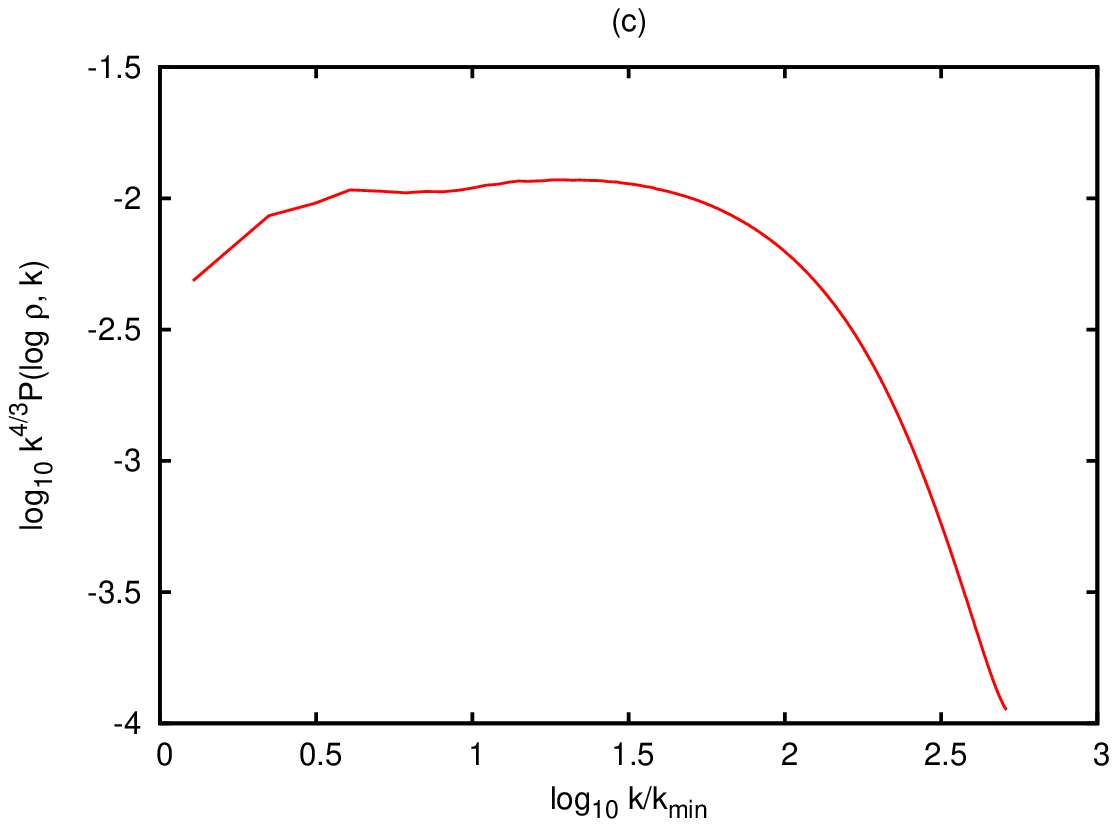}{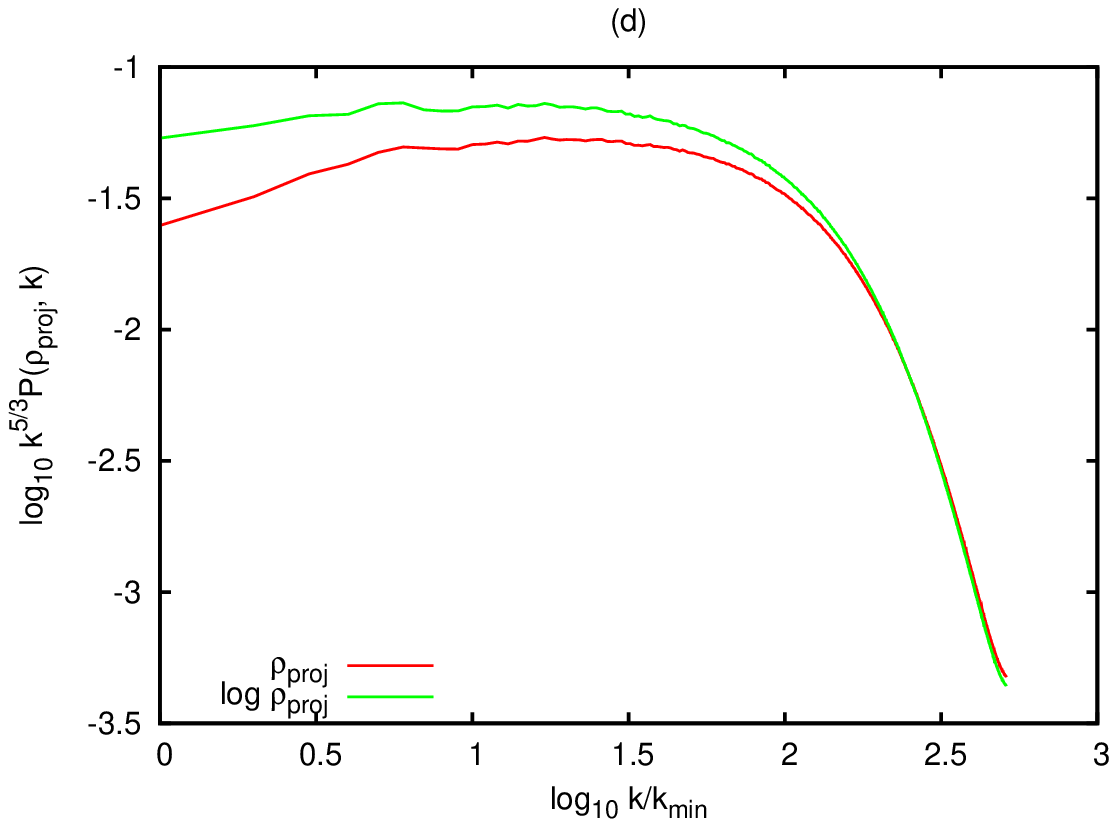}
\plottwo{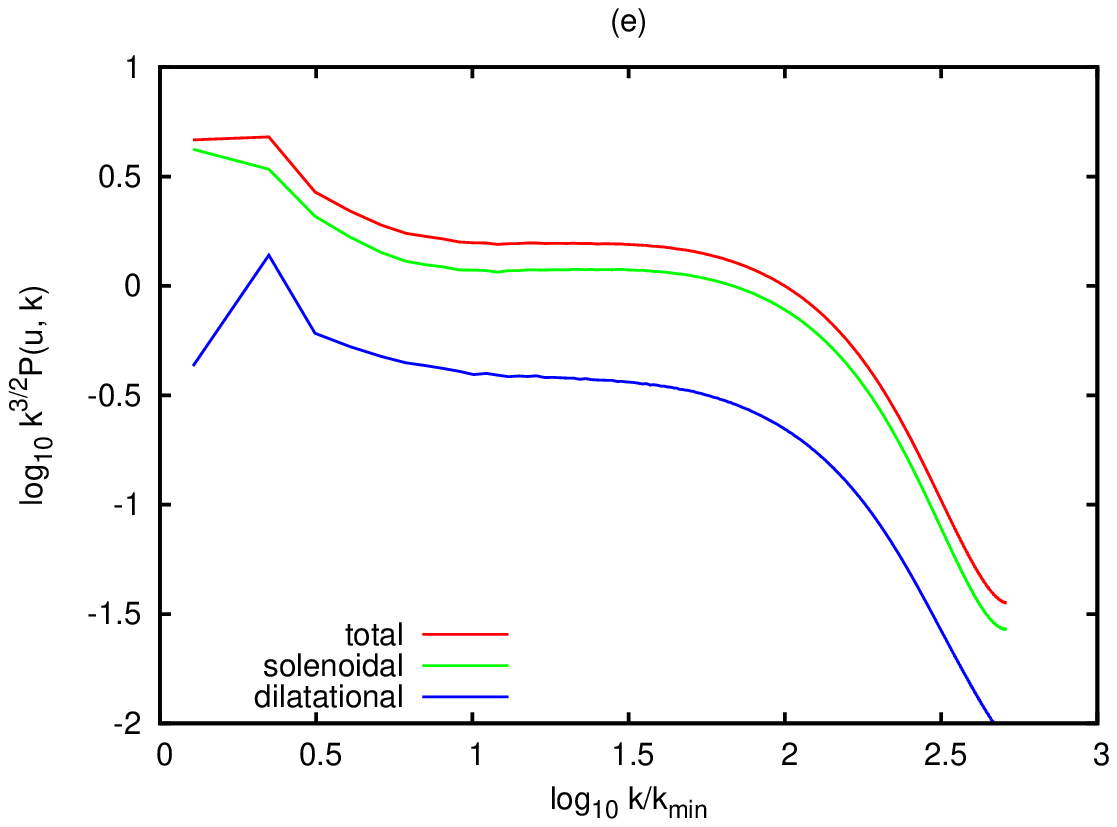}{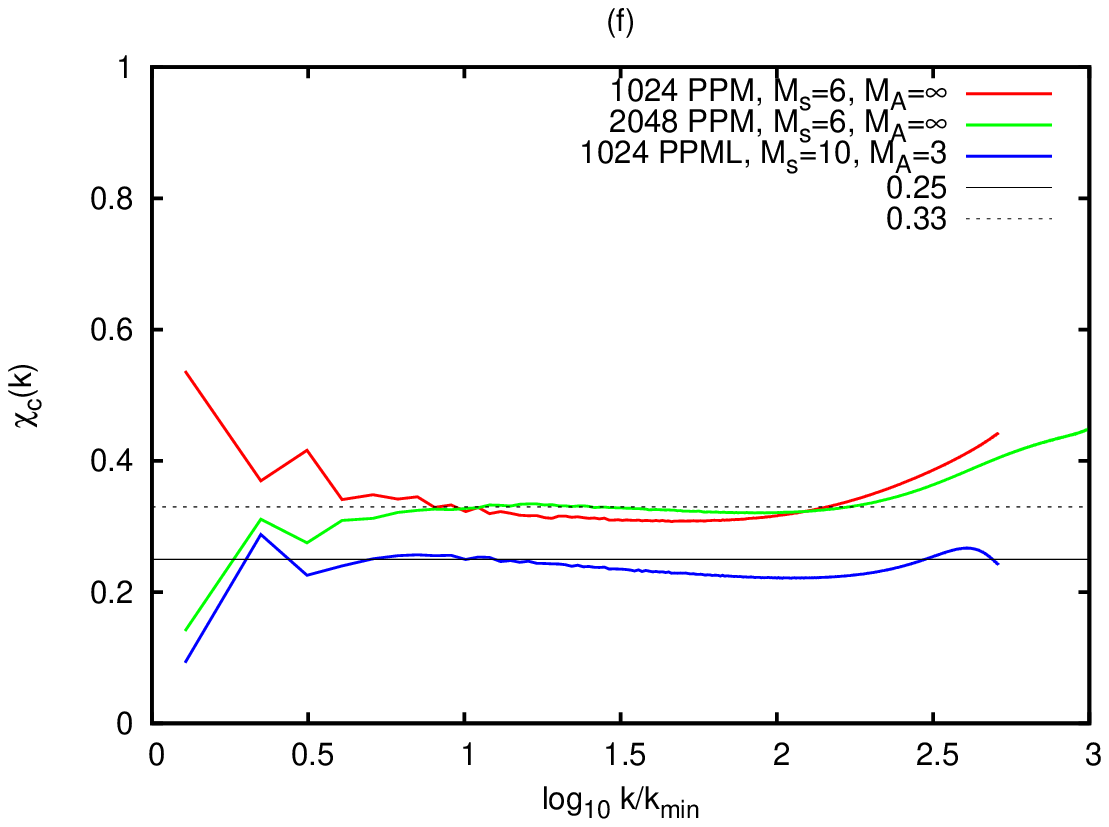}
\plottwo{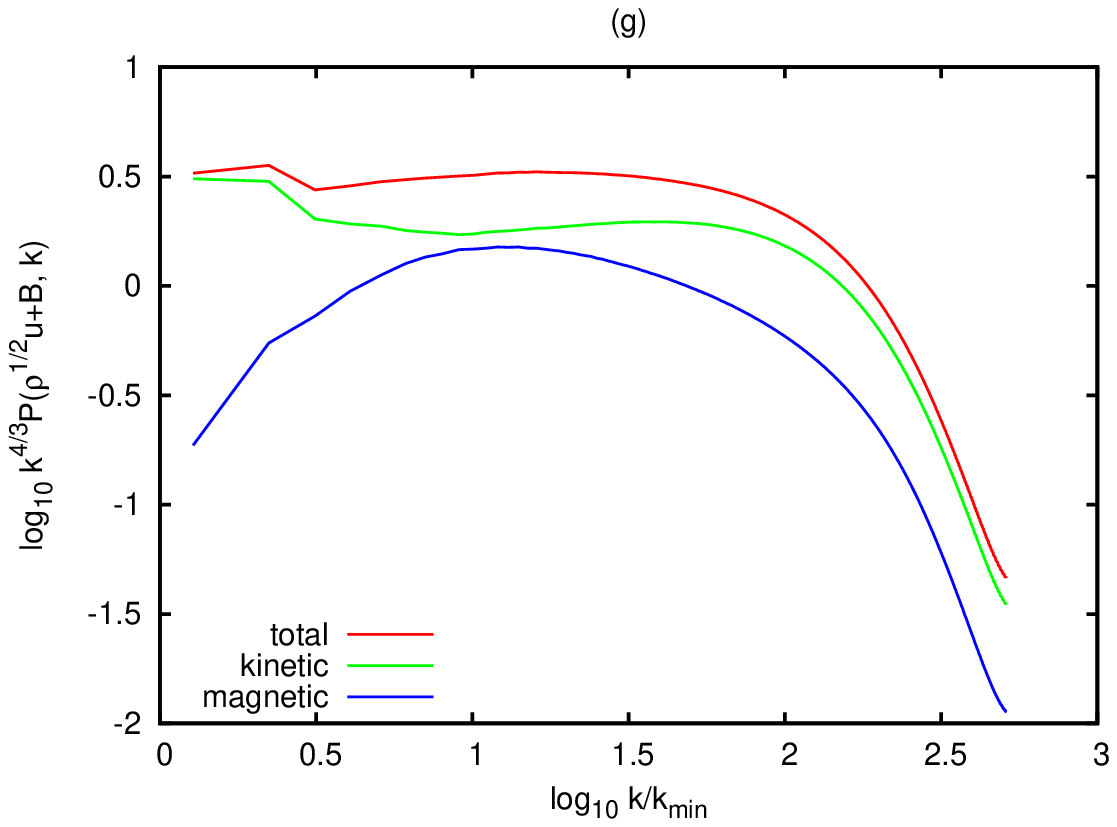}{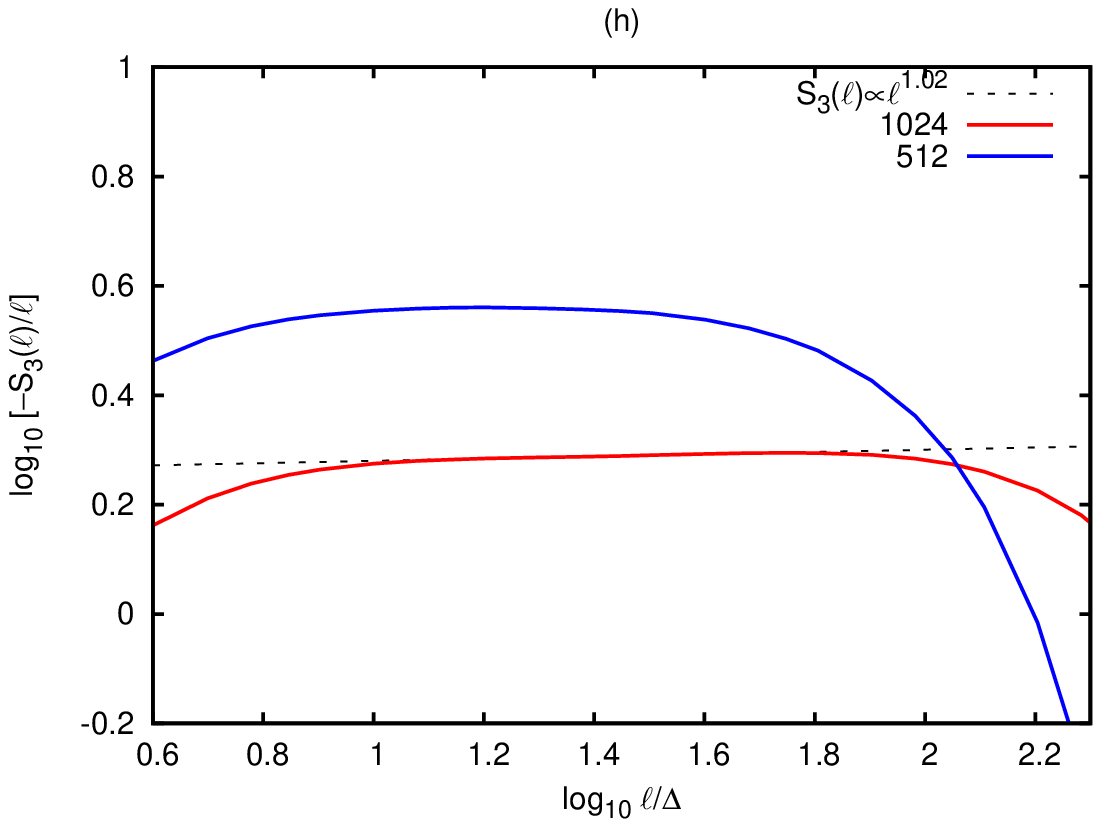}

\caption{\footnotesize Time-average statistics of MHD turbulence at $M_s=10$ and $\beta_0=2$, from PPML 
simulation at grid resolution $1024^3$ cells: 
({\bf a}) the gas density pdf and lognormal fits for $\log_{10}{\rho}\in[-1,2.8]$;
({\bf b}) compensated power spectrum of the gas density;
({\bf c}) as {\bf b}, but for the logarithm density;
({\bf d}) as {\bf b}, but for the projected density and for the logarithm of projected density;
({\bf e}) as {\bf b}, but for the velocity and its solenoidal and dilatational parts;
({\bf f}) ratio of dilatational to total velocity power $\chi_C$ as a function of wavenumber $k$;
({\bf g}) compensated power spectra of total, kinetic, and magnetic energy.
({\bf h}) compensated third-order structure functions $S_3(\ell)$ for generalized 
Els\"asser fields ${\bf Z}^{\pm}$ from $512^3$ and $1024^3$ simulations.
}
\label{figfour}
\end{figure}

The probability density function (pdf) of the gas density is one of the most important
inputs for the theories of star formation in molecular clouds as it determines the
amount of the dense material available for gravitational collapse to form stars 
\cite{padoan.02,krumholz.05,padoan....07}. The ability to predict the pdf is important in other extreme
astrophysical environments, such as, e.g., accretion onto ``seed'' black holes in
the first galaxies \cite{milosavljevic...09}. It is also known as one of the most 
robust statistical characteristics for compressible isothermal turbulence as its 
lognormal shape \cite{vazquez94,padoan..97,nordlund.99,biskamp03} gets quickly established in numerical 
simulations, e.g. \cite{kritsuk...07a}. Here we demonstrate the lognormal form of 
the pdf in our $\beta_0=2$ model which gives a perfect fit for a range of $\sim10^8$ 
in probability at high densities (Fig.~\ref{figfour}{\bf a}). Compared to the 
unmagnetized cases at $M_s=6$ studied in \cite{kritsuk...07a,kritsuk...09}, the 
lognormal distribution in the MHD case with $M_s=10$ is, as expected at higher Mach
numbers, somewhat wider. If we parametrize the standard deviation $\sigma$ as a 
function of the rms sonic Mach number, $\sigma^2\equiv\ln(1+b^2M_s^2)$, 
we get $b=0.22$ and 0.27 as the best fit values for $\log_{10}\rho\in[-1, 2.8]$ in 
the MHD and HD cases, respectively. The mean of the distrubution is also a function 
of the standard deviation, $\left<\ln\rho\right>=-\frac{1}{2}\sigma^2$. For more 
detail on the dependence of the pdf on the value of $\beta_0$ see \cite{kritsuk...09}.

The density power spectrum (Fig.~\ref{figfour}{\bf b}) scales roughly as $k^{-2/3}$,
shallower than the slope of $-1$ in the HD case at a smaller $M_s=6$ \cite{kritsuk...07a}.
The power spectrum of the logarithm of density (Fig.~\ref{figfour}{\bf c}) scales
somewhat steeper, $k^{-4/3}$, but still shallower than $-5/3$ found at $M_s=6$ 
\cite{kritsuk...09}. The power spectrum of the logarithm of projected gas density 
roughly follows the scaling $k^{-5/3}$, which is somewhat steeper than the scaling 
of the density power spectrum (Fig.~\ref{figfour}{\bf d}), but still shallower than 
$-2$ measured for the HD case at $M_s=6$ \cite{kritsuk...09}. The sensitivity of the 
density spectra to the value of $\beta_0$ is discussed in detail in \cite{kritsuk...09}.

The velocity power spectrum as well as the results of the Helmholtz decomposition of
the velocity ${\bf u}$ into the solenoidal (${\bf \nabla\cdot u}_s\equiv0$) and dilatational 
(${\bf \nabla}\times{\bf u}_c\equiv0$) parts, ${\bf u}={\bf u}_s+{\bf u}_c$, are shown in 
Fig.~\ref{figfour}{\bf e}. The velocity power index is very close to $-3/2$
at $k/k_{min}$ around 30, the value proposed for an isotropic incompressible case with 
turbulent energy equipartition \cite{iroshnikov63,kraichnan65}, and the spectrum is dominated by 
the solenoidal component which follows the same scaling. Note that in the HD simulation
of Ref.~\cite{kritsuk...07a} this range of wavenumbers is buried under the bottleneck
bump which is known to be much weaker, if exists at all, in the MHD models.
At lower wavenumbers (but still within the inertial range), the velocity spectrum tends 
to get steeper, as should be expected for a strongly supersonic regime. The spectrum for 
the dilatational component of velocity is generally slightly steeper, as was previously 
reported for non-magnetized turbulence \cite{porter..02,kritsuk...07a}. 
Figure~\ref{figfour}{\bf f} compares the fraction of dilatational component in the velocity 
power, $\chi_c(k)\equiv P({\bf u}_c,k)/P({\bf u},k)$,  from our PPML MHD model at 
$\beta_0=2$ with those in our 
previous PPM $M_s=6$ HD runs at resolutions of $1024^3$ and $2048^3$ cells 
\cite{kritsuk...07a,kritsuk...07b,kritsuk...09}. In both HD runs $\chi_c(k)\approx1/3$ in the
inertial range, even though we used different driving forces and different 
parameterizations for PPM diffusion in these simulations. The fact that $\chi_c(k)$ tends 
to be close to $1/3$ in isotropic flows at high Mach numbers can be explained by purely 
geometric considerations. At the same time, in our MHD model $\chi_c(k)\approx1/4$ 
or even somewhat lower. A lower mean level of saturation for $\chi_c(k)$ in magnetized flows  
at high Mach numbers can be explained by the additional magnetic pressure term in the momentum
conservation equation. A similar value for $\chi_c(k)$ was obtained in Ref.~\cite{lemaster.09} 
for a model with stronger field ($\beta_0=0.02$) at the same grid resolution, cf. \cite{boldyrev..02}. The 
$k^2$-compensated power spectra of dilatational velocities in Fig.~14 of Ref.~\cite{lemaster.09}, 
however, demonstrate an unusual rise of power near the Nyquist frequency that we do not
observe in our simulations with PPML. Neither do we see such rise in our HD models with PPM 
(Fig.~\ref{figfour}{\bf e}). The origin of this sharp rise in compressions at 
high wavenumbers in simulations with the {\em Athena} code \cite{stone....08} is unclear.

\subsection{Ideal Invariants\label{4.2}}
Since the total energy $E_T$ is a conserved quantity, it is instructive to follow its 
distribution as a function of scale, 
$E_T(k)\equiv\frac{1}{2}\left[ P(\rho^{1/2}{\bf u},k) + P({\bf B},k)\right]$,
see Fig.~\ref{figfour}{\bf g}. It appears that to a good approximation the total energy follows
a simple scaling law $E_T(k)\sim k^{4/3}$. It is hard to believe, though, that this scaling
exponent is universal. We have shown that in non-magnetized turbulence the power index for 
$P(\rho^{1/2}{\bf u},k)$ does depend on the rms Mach number of the flow \cite{kritsuk...07a}. 
In MHD models, the scaling of $E_T(k)$ would also depend on the saturation level of the 
magnetic energy. Nevertheless, it is interesting to see how a combination of the magnetic
and kinetic energy, each of which do not display a clear extended scaling range 
(Fig.~\ref{figfour}{\bf g}) add up to result in an extended flat stretch of the total
energy spectrum. Is this coincidental? Or is this a result of self-organization in
supersonic super-Alfv\'enic MHD turbulence?

A better candidate for universal scaling is the energy transfer rate. Indeed, our HD
simulations have shown that $P(\rho^{1/3}{\bf u},k)$ follows the Kolmogorov $-5/3$ scaling
at $M_s=6$ \cite{kritsuk...07a}.
There is a set of exact scaling laws for homogeneous and isotropic incompressible MHD 
turbulence analogous to the 4/5-law of Kolmogorov for ordinary turbulence in neutral fluids
\cite{chandrasekhar51,politano.98a,politano.98b}. 
The MHD laws can be expressed in terms of Els\"asser fields, 
${\bf z}^{\pm}\equiv{\bf u}\pm{\bf B}/\sqrt{4\pi\rho}$ \cite{elsasser50}, as
$S_{\parallel,3}^{\pm}\equiv\left<\delta z^{\mp}_{\parallel}(\ell) 
[\delta z_i^{\pm}(\ell)]^2\right>=-\frac{4}{d}\epsilon^{\pm}\ell$,
where $\delta{\bf z}_{\parallel}(\ell)\equiv[{\bf z}({\bf x+\hat{e}}\ell)-{\bf z}({\bf x})]\cdot{\bf \hat{e}}$, 
$d$ is the space dimension, ${\bf\hat{e}}$ is a unit vector with arbitrary direction, 
${\bf\hat{e}}\ell$ is the displacement vector, the brackets denote an ensemble average, 
and summation over repeated indices is implied. Equivalently, the scaling laws can be 
rewritten in terms of the basic fields (${\bf u}$, ${\bf B}$), but in MHD there are 
no separate {\em exact} laws for the velocity or the magnetic field alone. It is
important not to neglect the correlations between the ${\bf u}$ and ${\bf B}$ fields
or ${\bf z}^+$ and ${\bf z}^-$ fields constrained by the invariance properties of the 
equations in turbulent cascade models. In supersonic turbulence, it is equally important
to properly account for the density--velocity and density--magnetic field correlations.
In incompressible MHD, the total energy $E_T$ and the 
cross-helicity $H_C$ play the role of ideal invariants
and, thus, the (total) energy transfer rate $\epsilon^T=(\epsilon^++\epsilon^-)/2$ 
and the cross-helicity transfer rate $\epsilon^C=(\epsilon^+-\epsilon^-)/2$.
For vanishing magnetic field ${\bf B}$, one recovers the Kolmogorov 4/5-law, {\small
$\left<[\delta {\bf u}_{\parallel}(\ell)]^3\right>=-4/5\epsilon\ell$}, where $\epsilon$ is the mean
rate of kinetic energy transfer \cite{politano.98b}.

Numerical simulations generally confirm these incompressible scalings, although the 
Reynolds numbers are perhaps still too small to reproduce the asymptotic linear behavior 
with a desired precision
and the results are sensitive to statistical errors \cite{biskamp.00,porter..02,boldyrev..06}. 
Often, the absolute value of the longitudinal difference is used and still a linear scaling is
recovered numerically, while there is no rigorous result for the normalization constant in this 
case. The third-order transverse velocity structure functions also show a linear scaling in 
simulations of nonmagnetized turbulence \cite{porter..02,kritsuk...07a} and the difference
of scaling exponents for longitudinal and transverse structure functions can serve as a
robust measure of statistical uncertainty of the computed exponents \cite{kritsuk.04}.
In MHD simulations, the third-order structure functions of the Els\"asser fields 
$\left<|\delta z^{\mp}_{\parallel}(\ell)|^3\right>$ were shown
to have an approximately linear scaling too \cite{biskamp.00,momeni.08}, but again there 
is no reason for this result to universally hold in all situations since the correlations 
between the ${\bf z}^{\pm}$ fields play an important role in nonlinear transfer processes.

Can the 4/3-law for incompressible MHD turbulence be extended to supersonic regimes in 
molecular clouds? As we discussed in Ref.~\cite{kritsuk...09}, a proper density weighting 
of the velocity, $\rho^{1/3}{\bf u}$, preserves the approximately linear scaling of the 
third-order structure functions at high Mach numbers in the nonmagnetized case (for a more 
involved approach to density weighting see Ref.~\cite{pan..09}). It is straightforward to 
redefine the Els\"asser fields for compressible flows using this ``1/3-rule'', 
${\bf Z}^{\pm}\equiv\rho^{1/3}({\bf u}\pm{\bf B}/\sqrt{4\pi\rho})$,
so that they match the original ${\bf z}^{\pm}$ fields in the incompressible limit and reduce
to $\rho^{1/3}{\bf u}$ in the limit of vanishing ${\bf B}$. The new ${\bf Z}^{\pm}$ fields
can have universal scaling properties in homogeneous isotropic turbulent flows with a broad 
range of sonic and Alfv\'enic Mach numbers. 


We use data from our $1024^3$ MHD simulation 
to support the conjecture proposed in Ref.~\cite{kritsuk...09} and compute the third-order structure 
functions $S_{\parallel,3}^{\pm}$ and $S_{\perp,3}^{\pm}$ defined above using the ${\bf Z}^{\pm}$ 
fields. Since we are interested in the energy transfer through the inertial interval, we 
compute the sum of $S_{\parallel,3}^-$ and $S_{\parallel,3}^+$ which 
determines the energy transfer rate $\epsilon^T$ in the incompressible limit. To further 
reduce statistical errors, we combine the transverse and longitudinal structure functions, 
but the absolute value operator was not applied to the field differences. 
Figure~\ref{figfour}{\bf h} shows the compensated scaling for $S_3(\ell)\equiv(S_{\parallel,3}^-+
S_{\parallel,3}^+ + S_{\perp,3}^- + S_{\perp,3}^+)/4$ averaged over 100 flow snapshots and the best
linear fit $S_3(\ell)\sim\ell^{1.020\pm0.001}$ for $\log{\ell/\Delta}\in[1.1,1.8]$. For
comparison, we overplot data from our $512^3$ simulation discussed in 
Ref.~\cite{kritsuk...09}. Clearly, at $1024^3$, we get a more extended scaling 
range and confirm the linear scaling found in \cite{kritsuk...09}. Howefer, further work is
needed to verify this result with high resolution simulations at different levels of
magnetization.


\section{Conclusions}

We verified the applicability of the 1/3-rule of hydrodynamic supersonic turbulence to super-Alfv\'enic
regimes of MHD turbulence using data from a new simulation with the Piecewise Parabolic Method 
on a Local Stencil on a grid of $1024^3$ cells. 
Our results suggest that the energy transfer rate is approximately constant accross the
inertial interval and that the 4/3-law of incompressible MHD can be extended to the supersonic 
turbulent flows.

\ack
This research was supported in part by the National Science Foundation through 
grants AST-0607675 and AST-0808184, as well as through TeraGrid resources 
provided by NICS, PSC, SDSC, and TACC (MCA07S014 and MCA98N020).

\section*{References}

\providecommand{\newblock}{}

\end{document}